\documentclass[aps,prstper,reprint,showpacs,titlepage,longbibliography,floatfix]{revtex4-2}   

\usepackage[T1]{fontenc}	% should generally be included for better accented-word behavior
\usepackage[latin9]{inputenc}	% should generally be included for better accent behavior
\usepackage{geometry}		% for controlling page margins
\usepackage{graphicx}
\usepackage[above,below]{placeins}	% allows use of \FloatBarrier command to force section barriers
\usepackage{times}
\usepackage{makecell}

\usepackage{hyperref}  
\hypersetup{colorlinks=true,urlcolor=blue,citecolor=blue,linkcolor=blue}   
\usepackage{enumitem}          % package for handling list formatting
\setlist{nosep}                 % Tightest spacing for lists. `noitemsep` is more relaxed

\begin{document}

\begin{titlepage}

  \title{Using machine learning to understand physics graduate school admissions}

\author{Nicholas T. \surname{Young}}
\affiliation {Department of Physics and Astronomy, Michigan State University, East Lansing, Michigan 48824}
\affiliation {Department of Computational Mathematics, Science, and Engineering, Michigan State University, East Lansing, Michigan 48824}

\author{Marcos D. \surname{Caballero}}
\affiliation {Department of Physics and Astronomy, Michigan State University, East Lansing, Michigan 48824}
\affiliation {Center for Computing in Science Education \& Department of Physics, University of Oslo, N-0316 Oslo, Norway}
\affiliation {CREATE for STEM Institute, Michigan State University, East Lansing, Michigan 48824}

  % \keywords{}

  \begin{abstract}
    Among all of the first-year graduate students enrolled in doctoral-granting physics departments, the percentage of female and racial minority students has remained unchanged for the past 20 years. The current graduate program admissions process can create challenges for achieving diversity goals in physics. In this paper, we will investigate how the various aspects of a prospective student's application to a physics doctoral program affect the likelihood the applicant will be admitted. Admissions data was collected from a large, Midwestern public research university that has a decentralized admissions process and included applicants' undergraduate GPAs and institutions, research interests, and GRE scores. Because the collected data varied in scale, we used supervised machine learning algorithms to create models that predict who was admitted into the PhD program. We find that using only the applicant's undergraduate GPA and physics GRE score, we are able to predict with 75\% accuracy who will be admitted to the program.  \clearpage
  \end{abstract}
  %% Adding the `\clearpage` is the hack to make the title page.  In 2020, the proceedings is
  %% going to be double blind.  This change makes it so that we can programmatically remove the
  %% title page.  In the future, other blinding measures should be taken as well (for example,
  %% removing self-citations).  This is not needed in 2019.

  \maketitle
\end{titlepage}

\section{Introduction}
Despite other science, technology, engineering, and mathematics (STEM) fields becoming more diverse over the past few decades, physics has lagged behind with only 20\% of bachelor's degrees awarded to women and only 11\% awarded to racial minorities \cite{ivie_beyond_2018}. These numbers do not improve when considering graduate degrees where 20\% of doctoral degrees are granted to women and 7\% are granted to racial minorities \cite{ivie_beyond_2018}. While this underrepresentation has both enrollment and retention causes, this paper will focus on the factors that may affect enrollment in physics graduate programs.

When considering enrollment in physics PhD programs, prior work has found that minority students in physics are less likely to apply to programs if they feel that they will not be admitted based on low GPA or GRE scores, or lack of research experience \cite{cochran_identifying_2018}. Further, given that many graduate programs have application fees, financial concerns may prevent students from applying to graduate programs that they believe they will not be admitted to. Therefore, it is important to understand what matters when applying to physics graduate programs while acknowledging that many factors that are not easily quantifiable matter.

Previous research into graduate admissions in physics has tended to take a broad approach, characterizing the graduate admissions process across the United States, both for master's and PhD programs \cite{potvin_investigating_2017,chari_admissions_2019} or focusing on a specific subset of universities such as elite universities \cite{posselt_inside_2016}. These studies find that faculty consider numerical measures such as undergraduate GPA and GRE scores most important in the admissions process and have been conducted by either observing the admissions process or by surveying faculty about what they believe to be most important in the admissions process. More recent work in physics graduate admissions has explored applicant perceptions of the various components of the application \cite{chari_understanding_2019}. Missing in this analysis is an investigation of the actual applications of prospective physics graduate students. To our knowledge, there has only been one such study \cite{posselt_metrics_nodate}.

Given that applications to graduate programs consist of numerical data such as GPA and GRE scores, categorical data such as gender, race, and ethnicity, and open-ended data such as letters of recommendation and personal statements, graduate admissions is an ideal target for machine learning. Indeed, machine learning approaches to understanding graduate admissions have been employed in computer science to study self-reported admissions data \cite{gupta_will_2016} and to streamline the review process \cite{waters_grade:_2014}. Machine learning methods have also been employed more broadly in higher education admissions to predict which admitted students will accept an offer to attend a small liberal arts school \cite{lux_applications_2016,basu_predictive_2019} and to predict which students are likely to be admitted and to complete their MBA \cite{moore_expert_1998}.

The goal of this work to further the study of graduate admissions in physics by analyzing the applications using a machine learning approach. In addition, unlike other studies in physics graduate admissions, this work represents a case study of a single institution rather than a broad look at the graduate admissions landscape. However, since physics is regarded as a high consensus discipline, that is, there is large agreement about what counts as legitimate admissions practices \cite{posselt_disciplinary_2015}, we expect our results to generalize to similar doctoral programs. Future work will investigate the robustness of the findings presented here.

\section{Methods}

\subsection{Data}
The data used in this study comes from admissions data recorded from 512 domestic applicants to a physics and astronomy PhD program at a large, research-intensive, Midwestern public university from 2014-2017. Domestic and international applicants do not undergo the same review process and hence we only analyze applications from domestic students here. The admissions process is unique at this university in that the applications are not reviewed by a central committee but rather members of the subdiscipline in which the student expresses interest. For example, if the student intends to study high energy physics, only high energy physicists would review the application and decide whether to accept the student.  The data include the applicant's undergraduate institution and grade point average (GPA), their general and physics GRE scores, and their physics subdiscipline of interest. As laws of the state where this university is located prohibit giving preferential treatment to applicants based on gender, race, ethnicity, or national origin, demographics were not recorded by the physics and astronomy department and hence, are not available to us. Overall, 48\% of the domestic applicants were offered admission into the program.

\subsection{Describing Undergraduate Institutions}
Because the name of the undergraduate institution in itself does not provide useful information to an algorithm, we created new factors to describe characteristics of the institutions. To describe the overall institution, we classified each institution as public or private, whether it is a minority serving institution (MSI), the region of the country it is located in (such as Northeast, Southwest, etc.), and the Barron's selectivity of the institution, which describes how selective the undergraduate program is and serves as a proxy for prestige. Classifications for the first three categories were taken from the most recent Carnegie Rankings \cite{indiana_university_center_for_postsecondary_research_carnegie_nodate} while the Barron's classification came from Barron's \textit{Profiles of American Colleges}. Because the overall reputation of the applicant's undergraduate university may not describe the physics program at that university, we also included factors related to the physics program such as the highest physics degree offered at the university and the size of the undergraduate program and PhD program if applicable. The size of the undergraduate and PhD programs were determined by the number of graduates of the program in a typical year. The programs were then classified as small, medium-small, medium-large, or large based on which quartile they fell into.  We used the Roster of Physics Departments with Enrollment and Degree Data to collect this data \cite{nicholson_roster_2016}. All factors appearing in our model are shown in Table \ref{tab:factors} and include the scale of measurement.

\begin{table}[tp]
\caption{Variables used in our model and their scale of measurement}
\begin{tabular}{ll}
\hline \multicolumn{1}{c}{\textbf{Factor}} & \multicolumn{1}{c}{\textbf{Measurement Scale}} \\ \hline
Undergraduate GPA                                                              & Continuous        \\
Verbal GRE score                                                               & Continuous        \\
Quantitative GRE score                                                         & Continuous        \\
Written GRE score                                                              & Continuous        \\
Physics GRE score                                                              & Continuous        \\
Proposed research area                                                         & Categorical       \\
Application year                                                                           & Categorical       \\
Barron's selectivity                                                           & Categorical       \\
\makecell[l]{Region of applicant's \\undergraduate institution}                                            & Categorical       \\
\makecell[l]{Type of physics program \\ at applicant's undergraduate \\institution}                                   & Categorical       \\
\makecell[l]{Size of undergraduate physics \\program at applicant's \\undergraduate institution} & Categorical       \\
\makecell[l]{Size of doctoral physics program \\at applicant's undergraduate \\institution}      & Categorical       \\
\makecell[l]{Applicant attended a minority \\serving institution}                     & Binary            \\
Public or Private  & Binary \\
Output variable: admitted status & Binary \\
\hline
\hline
\end{tabular}
\label{tab:factors}
\end{table}

\subsection{Justifying our choices of institutional factors}
Prior work has documented university pedigree is often considered in the application process because institutional quality is assumed to be a proxy for student quality \cite{posselt_inside_2016, paxton_perceived_2003}. Here, we measure institutional quality by Barron's selectivity and public or private status, with the assumption that physics faculty view private universities as more prestigious than public universities. We include region of the applicant's undergraduate university to account for the fact that the institution being studied is a public university and may therefore show a preference for students from the surrounding region.

Prior work has also found faculty exhibit a tendency to admit students like themselves, though it is more common among academics who graduated from elite institutions \cite{posselt_inside_2016}. Therefore, it is not unreasonable to expect that faculty may prefer to admit students who followed similar paths as they did, meaning students from large, doctoral institutions may be more likely to be admitted than students from smaller institutions. Additionally, we use the size of the undergraduate and PhD programs as proxies for the perceived prestige of the physics department, assuming a more prestigious physics department attracts more students and hence graduates more students.

\subsection{Random Forest Model}
To analyze our data, we used the conditional inference forest algorithm, a variant of the random forest algorithm shown to be less biased when the data includes both continuous and categorical variables \cite{strobl_bias_2007} such as those used in our model (see table \ref{tab:factors}). Random forest models in general are ensembles of individual decision trees, which use binary splits of the input features in order to make a prediction. The predictions are then averaged over the individual trees to obtain the overall prediction of the random forest. 

While there are multiple metrics used to assess random forest and other machine learning models, two of the most common are the accuracy and the area under the curve (AUC). The accuracy is simply the proportion of correct predictions made by the model. To ensure that the accuracy isn't inflated by overtraining, only a fraction of the available data is used to construct the model while the rest is used to test the predictive power. It is this remaining data that is used to calculate the accuracy of the model. The AUC is defined as the area beneath the receiver operator curve of the model, which visualizes the false positive rate against the true positive rate and varies between 0.5 and 1, with values greater than 0.7 signifying a good model \cite{araujo_validation_2005}. This area describes the proportion of positive cases that are ranked above negative cases in the data set. For example, for our data, the AUC represents the proportion of all random pairs of admitted and not-admitted applicants in which the admitted applicant is classified as admitted and the not-admitted applicant is classified as not-admitted.

In addition to making predictions, the random forest algorithm can determine the importance of each feature to the model, referred to as the feature importance. We use the AUC feature importance \cite{janitza_auc-based_2013} as it is less biased when input features differ in scale (as do our factors listed in table \ref{tab:factors}) and when the predicted variable is not split evenly between the two outcomes. As the feature importances are not assumed to follow any statistical distribution, there is no simple way to apply the idea of statistical significance to feature importances. We therefore apply the recursive backward elimination technique described in \cite{diaz-uriarte_gene_2006} to determine which features are important and which are not. We will refer to the features determined to be important via this process as meaningful features. For more information about random forest models, biases, and feature importance measures, see Young et al. \cite{young_identifying_2019}.

We chose to apply a random forest model instead of a more traditional technique for classifying data such as logistic regression (as used by Attiyeh and Attiyeh \cite{attiyeh_testing_1997} and Posselt et al \cite{posselt_metrics_nodate} to study graduate admissions) due to these feature importances. As feature importances measure all factors on the same scale, that is how much they change the area under the curve, factors of otherwise different scales can be compared. This is in contrast to logistic regression where the odds ratio for a continuous variable would measure the change in odds for a unit increase in the variable while the odds ratio for a categorical or binary variable measures the change in odds relative to a reference group.

We used R \cite{r_core_team_r:_2018} for our analysis and the party package \cite{hothorn_survival_2006,strobl_bias_2007,strobl_conditional_2008} to create a conditional inference forest model. We used 500 trees to build our forest and 70\% of our data to train the model. We ran our model 30 times, randomly selecting 70\% of our data for training each time, and averaged the feature importances over runs so that the resulting distribution of individual feature importances would be approximately normal. As the conditional inference forest algorithm has routines built in to handle missing data \cite{hapfelmeier_new_2014}, applicants with missing information were not removed from the data set.

\begin{figure}[b]
  \includegraphics[width=.9\linewidth]{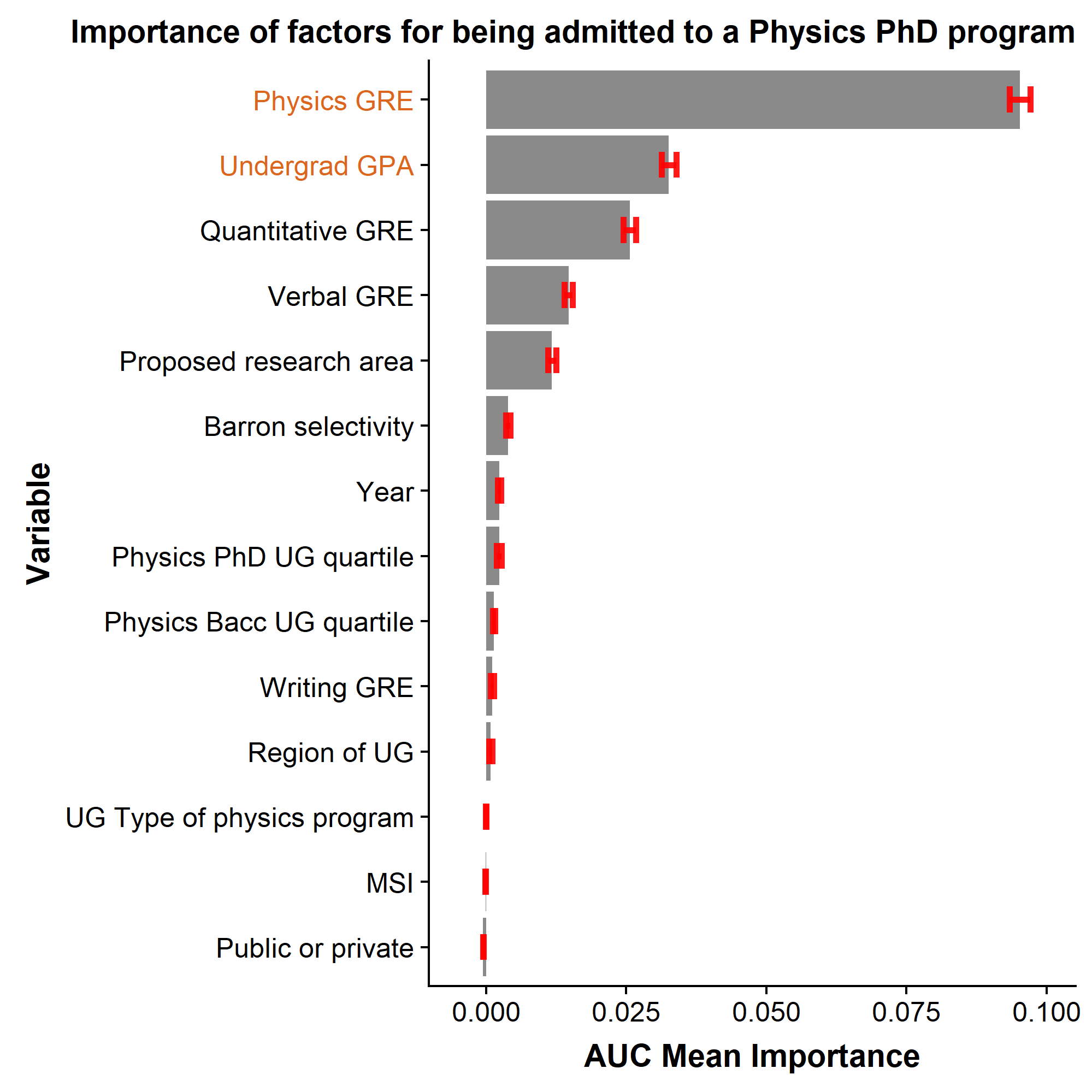}
  \caption{Averaged feature importances over 30 trials. Physics GRE score and undergraduate GPA, appearing in orange, were the only factors found to be meaningful.\label{fig1}}
\end{figure}

\section{Results}
Across the 30 runs, the average accuracy of our model predicting on the held-out data was $75.8\% \pm 0.6\%$ and the average AUC was $.821 \pm .002$. As our model's accuracy is significantly higher than the null accuracy of $52.3\%$, the percent of students who were not accepted, and our AUC is above 0.7, our model can be considered a reasonable model of the data. The accuracy and AUC were also stable with regard to how many trees we used to construct the model and the fraction of the data we used to train the model, with the accuracy changing by at most 1\% and the AUC by 0.005.

The feature importances averaged over the 30 runs are shown in Fig. \ref{fig1}. We find numerical factors such as the applicant's score on the physics GRE, the applicant's undergraduate GPA, the applicant's quantitative and verbal GRE scores, and their proposed research area to be more important in the application process than any factor describing the applicant's undergraduate institution. Using recursive backward elimination to determine the meaningful factors, we find the applicant's physics GRE score and their undergraduate GPA to be the only meaningful factors.

\begin{figure}[b]
  \includegraphics[width=.9\linewidth]{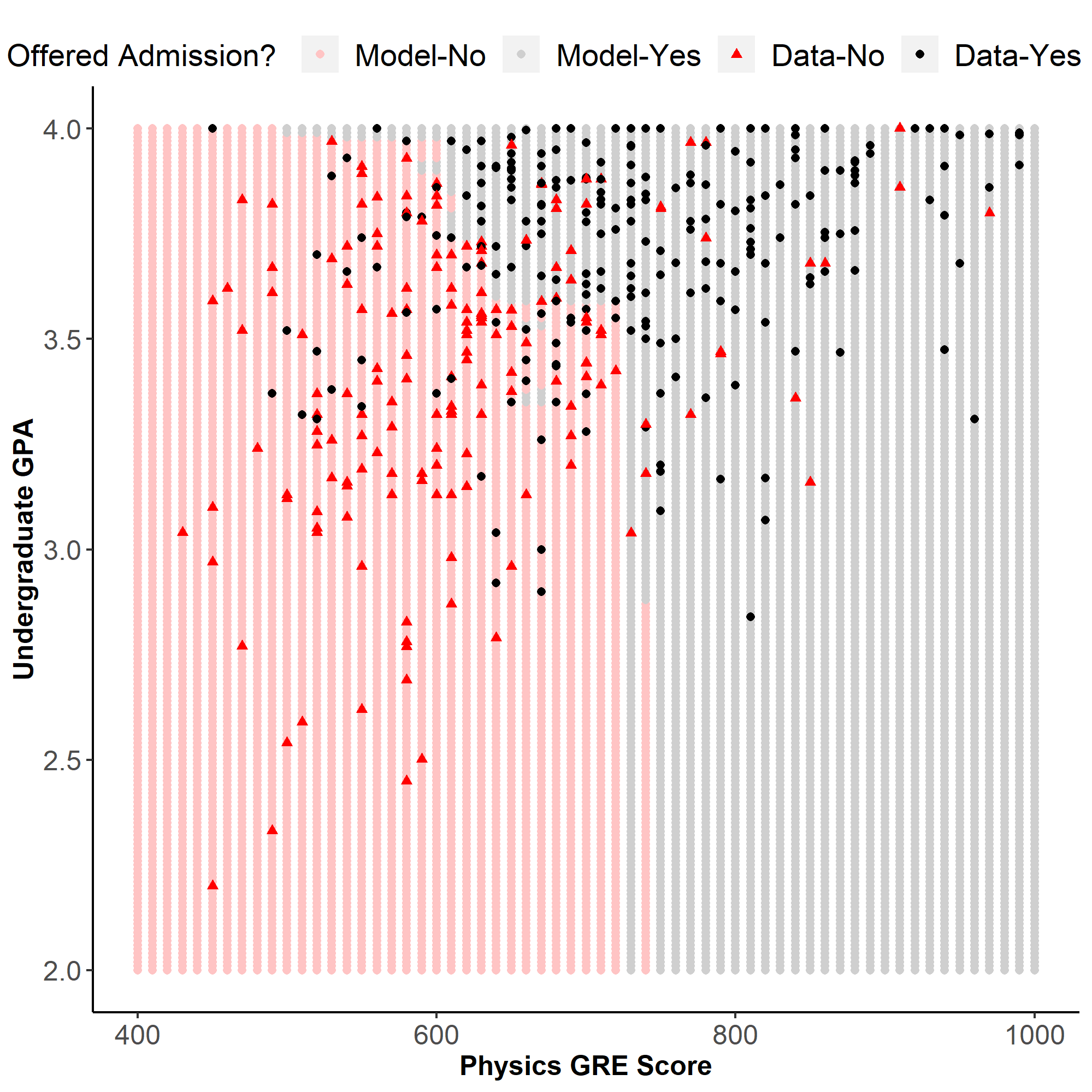}
  \caption{Plot of all applicant's physics GRE scores vs their undergraduate GPAs. The background coloring expresses the prediction of the model had an applicant had that score. \label{fig2}}
\end{figure}

To verify that the applicant's physics GRE score and their undergraduate GPA were indeed the only meaningful factors, we then reran our random forest model 30 times using only these two factors as the predictors. Our average accuracy was then $75.8\% \pm 0.6\%$ and our average area under the curve was $.817\pm .003$, which are not statistically different from the values we found using all fourteen factors shown in table \ref{tab:factors}. Further, if these factors are truly meaningful, then plotting an applicant's physics GRE score against their undergraduate GPA should show distinct groups. To test this idea, we generated all possible pairs of undergraduate GPA and physics GRE scores and ran them through our model to find the predicted admissions decision. The results are shown in Fig. \ref{fig2} with the data from all of the applicants in our study overlaid. While there does not appear to be a hard cutoff physics GRE score, there appears to be a rough cutoff around 700. Interestingly, the decision boundary has a negative slope between GPA of 3.6 and 4.0, suggesting a high GPA can partially compensate for a physics GRE score below the soft threshold.

\section{Discussion}
Perhaps unsurprisingly, we find numerical measures are the most important factors for determining whether a domestic applicant will be accepted into this physics graduate program, consistent with findings that graduate programs with a large number of applicants use numerical measures as a first pass to evaluate applicants \cite{posselt_inside_2016, potvin_investigating_2017}. While we find no evidence of a minimum physics GRE score, we do find evidence of a "rough cutoff" as described in Potvin et al. around 700. Nevertheless, some students who scored significantly above this threshold were not admitted. While we do not know the reasons why these students were not admitted, Posselt noted that faculty may not admit superior applicants if they do not believe the applicant will actually enroll in their program \cite{posselt_inside_2016}.

Overall, our findings of the meaningful factors for admission to a physics graduate program are consistent with Potvin et al.'s findings obtained by surveying physics graduate admissions directors. Notably, we also find that the physics GRE score, undergraduate GPA, quantitative GRE score, and proposed research area are more important than other factors while the undergraduate institution, GRE written score, and proximity/familiarity are less important factors. 

Interestingly, the program studied here appears to place more emphasis on the verbal GRE score than the average program. This may be because our study only looked at domestic students while Potvin et al.'s looked at all applicants. Because international students also take the TOEFL while domestic students do not and admissions directors ranked the TOEFL as more important the verbal GRE, the TOEFL may take the place of the verbal GRE and hence lower the perceived value of the verbal GRE relative to other factors.

Despite prior work suggesting institutional characteristics play an important role in graduate admissions, we did not find institutional or departmental characteristics to be meaningful to our model. Our result could be due to differences in methodology or due to institutional effects being influential but not dominant factors \cite{attiyeh_testing_1997}. Indeed, Posselt suggests institutional factors may be used to differentiate applicants with similar GPAs and GRE scores \cite{posselt_inside_2016}. Therefore, we may not have found institutional factors to be meaningful since they are used when primary factors such as GPA and physics GRE scores do not sufficiently separate applicants.

While we did not have access to other criterion included in the Potvin et al. such as application essays, research experiences, and recommendation letters, we were still able to create a model that correctly predicted whether an applicant would be admitted with 75\% accuracy based solely off the applicant's undergraduate GPA and physics GRE score. While undergraduate GPA is a significant predictor of completing a physics PhD, physics GRE is not, as those scoring near the top of the physics GRE only have a 7\% higher probability of completing their PhD than those scoring near the bottom \cite{miller_typical_2019}. As the GRE is not associated with completing a doctoral degree and is known to favor persons from majority groups in science \cite{miller_test_2014}, the outsized role of the GRE in the admissions process should be questioned. Indeed, the American Association of Physics Teachers recently released a recommendation against using the physics GRE in graduate admissions \cite{noauthor_statement_2019}.

\section{Limitations}
As with any model, there are a few limitations to our model. First, the data we used to make our model was not all the data that would be available to a faculty member evaluating an application. Therefore, it is possible that meaningful features other than GPA and physics GRE score could lie in the data that was unavailable to us.

Second, feature importances can be inflated when features used in the model are correlated, possibly causing some factors to be considered more important than they actually are. There do exist methods for controlling for correlations in random forest; however, unlike the conditional inference forest algorithm used here, those methods can only be used with data containing no missing values. Our initial analysis controlling for correlations suggests that undergraduate GPA and physics GRE scores are still the meaningful factors.

\section{Future Work and Conclusion}
Our work adds to the broader literature about graduate admissions and the process by which applicants are judged. Because minority students may not apply to graduate programs if they do not think they will be accepted, elucidating the factors that determine whether an applicant will be accepted is crucial. Simply increasing the number of applicants from underrepresented groups will not increase their representation unless corresponding efforts are made to admit these students. While our results align with prior work, these results represent only one institution and may not be representative of all United States physics PhD programs. Our future work will apply a similar analysis to other physics graduate programs to test the generalizability of our results. Given the unique structure of the admissions process at this university, graduate programs with a more traditional admissions process may assign different weights to the various parts of an application. 

Further, the university studied here has recently moved to a rubric-based admissions format, designed to take into account non-cognitive competencies and program fit in addition to the more traditional admissions criteria such as GPA and GRE scores. Our future work will examine how including these new criteria may change the factors that are most predictive of an applicant being admitted to the program.

\acknowledgments{We would like to thank the graduate program directors at the studied university for providing application data and for providing insight into their admissions process.}

% \bibliographystyle{apsrev} % supercedes the longbibliography option, so leave commented out if you want to display article titles
%\bibliography{Grad_school,Random_Forest} % don't include the .bib suffix
\bibliography{output}
\end{document}